\documentclass[runningheads]{llncs}
\usepackage{graphicx}
\usepackage{color}
\usepackage{listings}
\usepackage{tabularx} 

\newcommand{\opjump}{\textsf{JUMP}}
\newcommand{\opcreate}{\textsf{CREATE}}

\newcommand{\oprevert}{\textsf{REVERT}}

\newcommand{\opmystop}{\textsf{STOP}}
\newcommand{\opcall}{\textsf{CALL}}
\newcommand{\opdelegatecall}{\textsf{DELEGATECALL}}
\newcommand{\opcallcode}{\textsf{CALLCODE}}
\newcommand{\opjumpi}{\textsf{JUMPI}}

\newcommand{\reentrancy}{re-entrancy }
\newcommand{\createbase}{create-based re-entrancy }
\newcommand{\crossfunction}{cross-function re-entrancy }

\begin{document}
\title{Hunting for Re-Entrancy Attacks in Ethereum Smart Contracts via Static Analysis}
%

\author{Yuichiro Chinen\inst{1} \and 
Naoto Yanai\inst{1} \and
Jason Paul Cruz\inst{1}
\and \\ 
Shingo Okamura\inst{2}}
%
%
\institute{Osaka University, 1-5 Yamadaoka, Suita, Osaka, Japan 
\and
National Institute of Technology, Nara College, Nara 639-1080, Japan
}
\maketitle              
\begin{abstract}
Ethereum smart contracts are programs that are deployed and executed in a consensus-based blockchain managed by a peer-to-peer network. 
Several re-entrancy attacks that aim to steal Ether, the cryptocurrency used in Ethereum, stored in deployed smart contracts have been found in the recent years. A countermeasure to such attacks is based on dynamic analysis that executes the smart contracts themselves, but it requires the spending of Ether and knowledge of attack patterns for analysis in advance. 
In this paper, we present a static analysis tool named \textit{RA (Re-entrancy Analyzer)}, a combination of symbolic execution and equivalence checking by a satisfiability modulo theories solver 
to analyze smart contract vulnerabilities to re-entrancy attacks. 
In contrast to existing tools, RA supports analysis of inter-contract behaviors by using only the Etherum Virtual Machine bytecodes of target smart contracts, i.e., even without prior knowledge of attack patterns and 
without spending Ether. Furthermore, RA can verify existence of vulnerabilities to re-entrancy attacks without execution of smart contracts and it does not provide false positives and false negatives. 
We also present an implementation of RA to evaluate its performance in analyzing the vulnerability of deployed smart contracts to re-entrancy attacks and show that RA can precisely determine which smart contracts are vulnerable.
\end{abstract}
\section{Introduction} \label{Introduction}


\textbf{Ethereum.} 
Ethereum, which is often described as ``the world computer,'' is a global, open-source platform for decentralized applications and execution of programs called \emph{smart contracts}, which are software programs recorded on the Ethereum blockchain\footnote{Hereafter, ``contract'' and ``smart contract'' are used interchangeably but have the same meaning. 
Likewise, ``the blockchain'' will be used to refer to the Ethereum blockchain, unless otherwise specified.} and executed by the \emph{Ethereum Virtual Machine (EVM)}. The EVM is a virtual machine that runs codes called \emph{EVM bytecodes} and is the runtime environment for smart contracts in Ethereum. Ethereum enables developers to build smart contracts with built-in functions and gain the benefits of cryptocurrency and blockchain technologies.

\textbf{Research Motivation.} 
The blockchain is decentralized and transparent by nature, and thus anyone can read the bytecodes of deployed contracts. Moreover, smart contracts typically contain financially valuable data and therefore create a criminogenic environment for adversaries. 
The attack on ``The DAO'' on June 2016 is the most infamous case of an attack on smart contracts. In the attack, a vulnerability called \textit{re-entrancy}, where the main contract calls an external contract which again calls into the calling contract within a single transaction, was utilized to steal more than 60 million US Dollars worth of Ether. 
With the transparency of the Ethereum blockchain, the vulnerabilities of deployed contracts can be utilized permanently as a springboard to attacks.
Indeed, many attacks~\cite{atzei2017survey} and honeypots~\cite{torres2019art} have been found on Ethereum in the past years.

\textbf{Security Analysis of Smart Contracts.} 
Based on these backgrounds, program analysis for the security of Ethereum smart contracts is an urgent and significant research theme. 
In this paper, we focus on the analysis of EVM bytecodes of smart contracts instead of their corresponding source codes. 
Analysis of bytecodes brings a number of advantages: (i) the analysis is independent of a high-level language that is periodically updated, and (ii) bytecodes can be obtained directly from the blockchain even if the corresponding source codes are unpublished. 
Using static analysis, analysts can easily judge whether a deployed contract has benign or malicious codes, e.g., vulnerable or honeypots.

In early literature on analysis of EVM bytecodes, \textit{formal verification}~\cite{bhargavan2016formal,hildenbrandt2018kevm,tsankov2018securify,kalra2018zeus,grishchenko2018semantic} is a leading approach for verifying a specification of bytecodes and \textit{symbolic execution}~\cite{luu2016making,nikolic2018finding,torres2018osiris,liu2018s,chen2017under} is used for exploring bytecodes in a depth-first search fashion by extracting control flow graphs (CFGs). 
To the best of our knowledge, formal verification hovers at a level of just an abstraction of Ethereum bytecodes despite providing precise verification in general. In other words, 
using formal verification to analyze vulnerabilities in their entirety is difficult. Moreover, according to Weiss et al.~\cite{Weiss2019}, rigid formal verification is often dismissed as it puts too high demands on analysts who are not experts in formal verification. 
On the other hand, symbolic execution can potentially support analysis of vulnerabilities via extraction of CFGs. Several prior works~\cite{luu2016making,nikolic2018finding,torres2018osiris} succeeded in analyzing some vulnerabilities, such as to re-entrancy attacks. 
However, symbolic execution only outputs CFGs from programs to be analyzed and does not support detection of the vulnerabilities themselves. 
Accordingly, analysis becomes often heuristic if an infeasible path exists in the programs. In particular, analysis results may produce many false positives and false negatives because of an infeasible path~\cite{chang2019scompile}. 


A recent work~\cite{rodler2018sereum} has found \reentrancy attacks that undermine
existing analysis tools~\cite{tsankov2018securify,kalra2018zeus,luu2016making} by creating a new contract or calling a different function via an external contract. 
Although a monitoring tool was also proposed in the same paper as a countermeasure against those attacks, the tool performs monitoring based on \textit{dynamic analysis} and is therefore unable to detect vulnerabilities unless the attacks occur during program execution. 
Accordingly, analysts need to implement and execute attack patterns by themselves in advance to check for vulnerabilities. 
Although generic attack patterns~\cite{ferreira2020aegis} have been presented formally, dynamic analysis remains impractical for analysis of smart contracts because it often requires analysts to know how an attack is launched as well as to spend Ether for the execution of contracts. 
Ideally, contracts should be analyzed based on only their bytecodes, i.e., with the use of \textit{static analysis}. 

\textbf{Research Goal.} 
This paper aims to \textit{design an inter-contract static analysis tool that (1)~uses only EVM bytecodes as input, (2)~eliminates false negatives and false positives, and (3)~does not require analysts to have a priori knowledge of the attacks on contracts.} 
For this goal, we focus on analysis of \reentrancy attacks~\cite{rodler2018sereum}. 
As evident in the attack on ``The DAO", \reentrancy attacks have a significant impact and many contracts have been found to be vulnerable to these attacks~\cite{kalra2018zeus,luu2016making}. 
Considering findings on new attacks~\cite{rodler2018sereum} in the recent years, our research goal can prove to be important and useful. 
Moreover, the design of a static analysis tool that is effective against new attacks is an open challenge~\cite{rodler2018sereum}.

\textbf{Contributions.} 
In this paper, we present a new static analysis tool named \textit{Re-entrancy Analyzer (RA)} for analyzing EVM bytecodes of smart contracts. RA can analyze re-entrancy vulnerabilities via inter-contract flows, for instance, by creating a new contract and calling a different function in the main contract via an external contract. 
RA does not require analysts to have prior knowledge of the attack patterns and pay Ether for analysis, and it does not provide false positives and false negatives. 
These advantages are achieved via integration of symbolic execution and equivalence checking with a satisfiability modulo theories (SMT) solver.
Furthermore, we provide an implementation of RA and evaluate it by utilizing publicly available reference implementations of \reentrancy vulnerabilities~\cite{rodler2018sereum,solidityDoc,knownattacks}. Our results confirm that, unlike Oyente~\cite{luu2016making}, 
RA can analyze precisely the vulnerabilities of deployed contracts against state-of-the-art re-entrancy attacks~\cite{rodler2018sereum}. 
We have released the source codes of RA via GitHub (\url{https://github.com/wanidon/RA}) for reproducibility and as reference for future works.



Our main contribution is the creation of a new symbolic execution method named \textit{symbolic re-entrancy emulation}, which emulates re-entrancy attacks by connecting different contracts with each other. 
As will be described in detail in Section~\ref{Problem Statement}, Oyente and its extensions~\cite{tsankov2018securify,kalra2018zeus} 
do not support inter-contract analysis, for example, CFGs become fragments.  
On the other hand, we developed a module that localizes stored data on the blockchain in each execution path and a module that stacks a return address of an account information for each path.
Using these modules, RA can support inter-contract analysis and emulate the behavior of re-entrancy attacks in a symbolic fashion via an internal implementation of a dummy contract, i.e., a contract that executes other contracts. 

As another important contribution, 
we developed a new method named \textit{vulnerability verification}, which verifies vulnerabilities by utilizing the Z3 SMT solver in the CFGs obtained from the symbolic re-entrancy emulation. In particular, on path conditions of the obtained CFGs, our method verifies whether program behavior on paths for executions with re-entrancy attacks is identical to that without the attacks, i.e., behavior on normal executions.
Using the methods above, RA can completely eliminate false positives and false negatives (See Section~\ref{Design of RA} for details).

\section{Technical Background} \label{Technical Background}

\textbf{Ethereum Smart Contracts and EVM.} 
In Ethereum, there are two kinds of accounts, namely, an externally owned account (EOA) and a contract account. EOAs have a private key that can be used to access the corresponding Ether or contracts. A contract account has smart contract code, which an EOA cannot have, and it does not have a private key. Instead, it is owned and controlled by the logic of its smart contract code.
In Ethereum, a smart contract refers to an \emph{immutable} computer program 
that is deployed on the blockchain and runs \emph{deterministically} in the context of the EVM. The immutability property indicates that, similar to any data published on a general blockchain, smart contract codes can be considered as trustworthy, i.e., once deployed, they cannot be changed or deleted. The deterministic property indicates that the execution of the coded functions of smart contracts will produce the same result for anyone who runs them.
Once deployed on the blockchain, a contract is self-enforcing and managed by the peers in the network, i.e., its functions are executed when the conditions in the contract are met. A smart contract is given an identity in terms of a contract address. Using this address, it can receive Ether and its functions can be executed. A contract is invoked when its contract address is the destination of a transaction, which is a signed message originating from an EOA, transmitted by the network, and recorded on the blockchain. Such transaction causes a contract to run in the EVM using the transaction (and transaction’s data) as input. The data indicate which specific function in the contract to run and what parameters to pass to that function. To incentivize peers to execute contract functions, Ethereum relies on \emph{gas}, which is paid in Ether, to “fuel computations”. The amount of gas needed to execute a transaction is relative to the complexity of the computations, thus also preventing infinite loops. 

Smart contracts are typically written in a high-level language such as Solidity~\cite{solidityDoc}. 
The source code is then compiled to low-level bytecode that runs in the EVM. 
The EVM is a simple stack-based architecture. Its instruction set is kept minimal to avoid incorrect implementations that could cause consensus problems.
The EVM is a global singleton, i.e., it operates like a global, single-instance computer that runs in all peers in the network. Each peer runs a local copy of the EVM to validate the execution of contract functions, and the processed transactions and smart contracts are recorded on the blockchain. 


\textbf{Static Analysis of Programs.} 
This paper focuses on the use of static analysis composed of CFGs, SMT solver, and symbolic execution. 
A CFG represents feasible paths of a program as a graph and is utilized for optimization by a compiler and static analysis of programs. Paths in a manner of sequential execution without branches are called \textit{basic blocks} and are identical to nodes of a CFG. 
Likewise, paths which are feasible via branches or jumps are represented by edges to connect with nodes. 

An SMT solver is a tool used for SMT problems. 
In contrast to satisfiability problems represented by proportional logic, SMT problems are represented by the first-order predicate logic which is more representative. 
By describing a specification to be verified in some logic formally, an SMT solver verifies whether a program satisfies the given specification. 

Symbolic execution is a method that pseudo-executes a program by replacing information unspecified from the program itself with symbolic values to represent any value. 
Symbolic execution is composed of CFGs and an SMT solver, and is suitable for analysis of smart contracts given that smart contracts utilize information on blockchains which are outside of program codes. 
Specifically, a condition to execute a path is called a \textit{path condition}. 
Path conditions at the beginning of program execution are valid, and a restriction is newly added to the path conditions when a branch occurs. 
In a case where a path condition contains a symbolic value, executing either one or both paths according to a condition, i.e., the condition is satisfied or not, is decided by checking the satisfiability of the condition.
Path conditions are often represented by first-order predicate logic, and their satisfiability is decided by an SMT solver described above. 

\section{Motivating Example and Technical Difficulties} \label{Problem Statement}

In this section, we recall the fallback function and the re-entrancy attacks shown by Rodler et al.~\cite{rodler2018sereum} as our motivating example and then discuss the technical difficulties in the analysis of the attacks.

\subsection{Re-entrancy Problem}

\textbf{Fallback Function.} 
Functions in Solidity are similar to classes in object-oriented languages. There are four types of Solidity functions, namely, \texttt{external}, \texttt{internal}, \texttt{public}, and \texttt{private}. A contract can have exactly one unnamed function called a \emph{fallback function}, which cannot have any arguments, cannot return anything, and should have external visibility. The fallback function of a contract is executed whenever the contract receives Ether without any data included. 
To receive Ether and add it to the total balance of the contract, the fallback function must be marked \texttt{payable}. If the contract does not have a fallback function, then it cannot receive Ether through regular transactions and throws an exception. In other words, if a contract is intended to not receive Ether, then the payable in the fallback function can simply be removed.
The fallback function is also triggered if someone tries to call a function that does not exist in the contract, and is often utilized for \reentrancy attacks.

\textbf{Create-Based Re-Entrancy Attack.} 
\opcreate\  is an instruction that creates a new contract during execution of a contract (we call this the original contract for convenience).
The new contract consists of initialization codes and the codes of its functions, and these codes are allocated by following a \opmystop\  instruction of the original contract. 
Once the initialization codes are executed, data is initialized and then bytecodes of the new contract are returned. 
These transitions via the initialization codes can be viewed as function calls by \opcreate. 
Whenever a new contract is created, its constructor will be executed immediately. 
In the attacks by Rodler et al.~\cite{rodler2018sereum}, 
a newly generated contract by \opcreate\ can issue further calls in its constructor to other contracts, including malicious contracts, via \opcall\ in the initialization codes. 
Here, the victim contract first creates a new contract and then updates its internal state. The newly created contract then calls a contract owned by an adversary. 
Consequently, the adversary maliciously withdraws from the victim contract via the re-entrancy.

\textbf{Cross-Function Re-Entrancy Attack.} 
Whenever function calls such that Ether is sent to an external contract account are executed, transactions are created by the \opcall\  instruction. 
In doing so, some data are sent together with the transaction. 
The most significant four bytes are a function ID of a caller function, which is obtained from a function name and a type name of variables. 
The callee contract obtains the function ID of the caller contract from the received transaction and then decides a function to be executed by checking if the ID is identical to that owned by the callee contract. 
If the callee contract does not own a function ID specified by a transaction, a fallback function is executed as described above. 
Cross-function re-entrancy attacks proposed by Rodler et al.~\cite{rodler2018sereum} are launched over multiple functions of the victim contract. 
Specifically, compared to classical re-entrancy attacks that are launched by re-entering the same contract via the same function, cross-function re-entrancy attacks are launched by re-entering the same contract via a \textit{different} function.

\subsection{Technical Difficulties} \label{Technical Difficulty}

We present some technical issues that need to be solved for the analysis of the attacks described in the previous section. 
First, existing tools~\cite{tsankov2018securify,kalra2018zeus,luu2016making} do not provide support for generating running states of a callee contract via opcodes such as \opcreate\ and \opcall, which utilize an external contract. Consequently, in these tools, a caller contract is unidentified uniquely from the standpoint of a callee contract and vice versa. 
More concretely, in the reference implementation of \reentrancy attacks~\cite{rodler2018sereum}, 
an initialization code for the \createbase attack is executed in the aforementioned manner. 
In doing so, the initialization code is on an \textit{infeasible} path for existing tools. 
Moreover, because the main body of the created code is determined by a return value from the initialization code, the contract by \opcreate\ is infeasible unless the initialization code is analyzed. Thus, the behavior of the created contract including \opcall\  is unknown during offline analysis. 
Likewise, when an external contract is called by \opcall\  for the cross-function re-entrancy attacks~\cite{rodler2018sereum}, 
analysis should be executed independently for each contract because running states of a callee contract are not generated. 
Consequently, utilizing path conditions for a caller contract is no longer meaningful to analyze behavior of a callee contract. 
Furthermore, there are many candidates of function combinations,
and thus analyzing vulnerability to \crossfunction attacks becomes difficult due to the potential state explosion~\cite{rodler2018sereum}. 

Second, a method that evaluates vulnerabilities should be considered as well. 
Several analysis tools~\cite{luu2016making,nikolic2018finding,chang2019scompile} report feasible paths but do not provide evaluation and detection of vulnerabilities. 
Accordingly, analysts often need to determine if a contract is vulnerable in a heuristic fashion, and hence many false positives and false negatives are produced. To eliminate false positives and false negatives, analysis tools should include a method that can evaluate vulnerabilities without requiring analysts to have prior knowledge of the attacks. 

\section{Design of RA}  \label{Design of RA}

In this section, we present \textit{Re-entrancy Analyzer (RA)}, a new static analysis tool for re-entrancy attacks on Ethereum smart contracts. 
We first describe our design concept and then describe symbolic re-entrancy emulation and vulnerability verification as main processes, including their implementation.

\subsection{Design Concept}


The main idea of RA is to combine symbolic execution and an SMT solver in a \textit{dual way}, i.e., \textit{symbolic re-entrancy emulation} and \textit{vulnerability verification}. 

First, modules that \textit{emulate} \reentrancy attacks based on symbolic execution are developed. 
Loosely speaking, by using an SMT solver to verify if conditions for calling to an external contract and generating basic blocks are satisfied, RA can identify which function is called. Consequently, RA can find path conditions via symbolic execution, which transits executions to each basic block recursively. 
Second, by utilizing the path conditions obtained from the emulation process above, RA \textit{verifies} if a resultant running state of a function call based on a fallback function is equivalent with the original behavior, i.e., without the fallback function, on the CFG obtained from the emulation. 
The verification is done by the SMT solver. 
While symbolic executions in literature~\cite{tsankov2018securify,kalra2018zeus,luu2016making} report only program behavior via CFGs obtained within a \textit{single} contract, RA emulates re-entrancy attacks including \textit{inter-contract} behavior and then it verifies the vulnerability of these contracts to re-entrancy attacks. Consequently, in contrast to the early literature, while the coverage of the analysis for re-entrancy attacks is improved drastically, RA can analyze the vulnerabilities precisely, i.e., without false positives and false negatives and without requiring analysts to have prior knowledge of attack patterns, even for state-of-the-art \reentrancy attacks~\cite{rodler2018sereum}.

\subsection{Tool Overview}

\begin{figure}[t]
        \begin{center}
            \includegraphics[width=\textwidth,height=6cm]{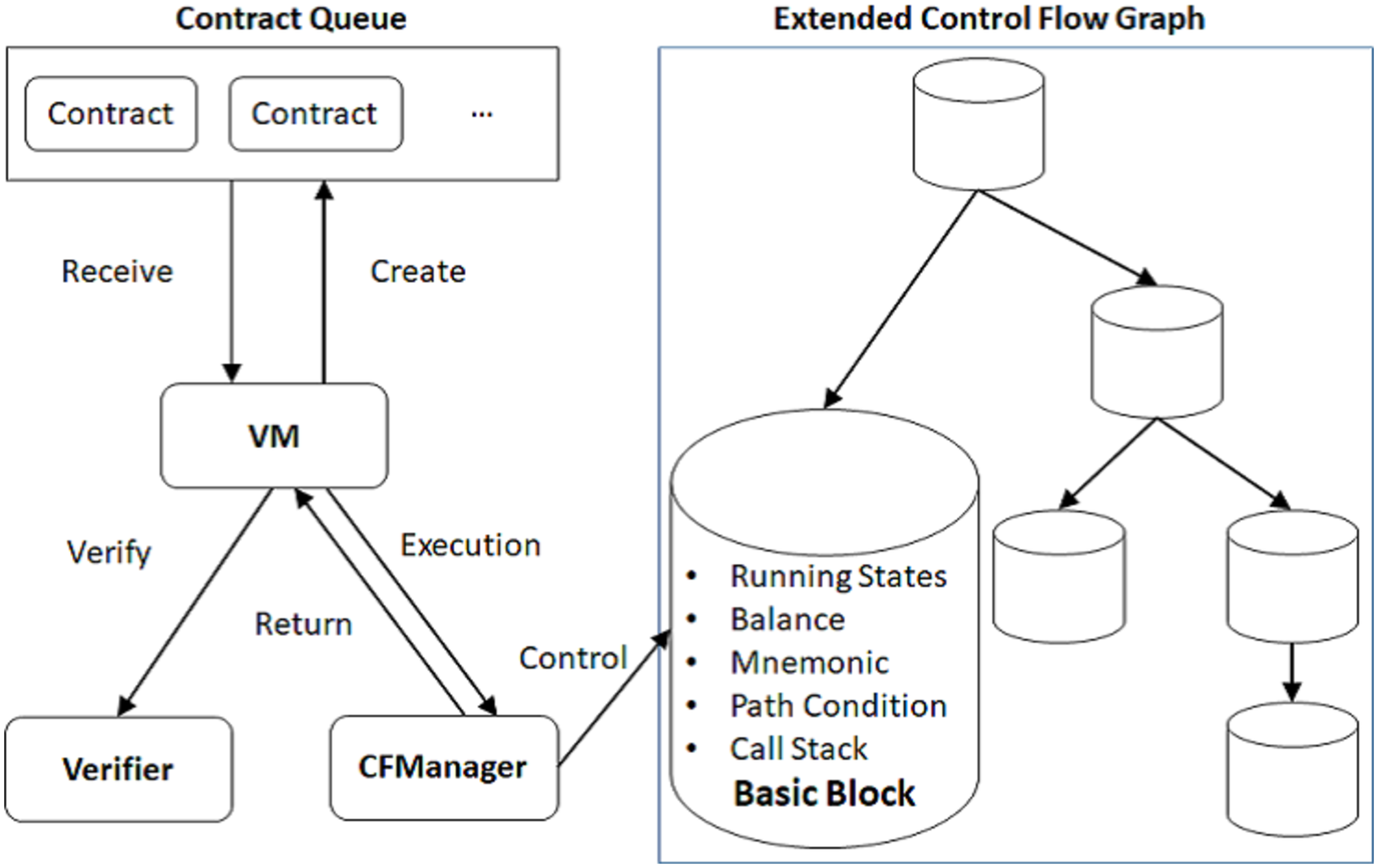}
        \end{center}
        \caption{Overview of RA}
        \label{architecture}
    \end{figure}

The overview of RA is as follows. 
First, for the symbolic re-entrancy emulation, RA generates CFGs from bytecodes whereby a newly created/called contract is represented within a flow of the main contract via symbolic execution. 
Then, for vulnerability verification, RA verifies whether the bytecodes are vulnerable to \reentrancy attacks by utilizing the Z3 SMT solver in accordance with path conditions obtained from the symbolic re-entrancy emulation. 
To do this, the following notions are defined for RA: 
\begin{itemize}
    \item \textit{Local-world state} is owned by a basic block in local and stores global information, such as balance or storage. 
    
    \item \textit{Call stack} is a stack that stores a return address to a basic block. 
    
    \item \textit{Contract queue} is a queue that stores bytecodes of a contract to be analyzed.  
\end{itemize}
An overview of RA including the notions is presented in Fig.~\ref{architecture}. 
RA mainly consists of three modules, namely, \textit{CFManager}, \textit{VM}, and \textit{Verifier}. 
The symbolic re-entrancy emulation process is mainly conducted by  CFManager and VM. On the other hand, the vulnerability verification process is executed by Verifier assisted by VM. 
We describe the role of each module below. 

CFManager handles feasible paths by directly operating data structures in a basic block for a CFG. A basic block contains the local-world state, the call stack, as well as mnemonic of instructions, running states, and path conditions. 
These give us information retrieved from the blockchain and return values, i.e., information obtained as a result of each caller/callee contract. Consequently, symbolic execution of each block can be covered even through an external contract. 
Then, VM receives a contract to be analyzed from the contract queue and then executes instructions in accordance with basic blocks via CFManager and checking conditions by Verifier. 
Finally, Verifier verifies existence of re-entrancy vulnerability by utilizing the Z3 SMT solver with the information obtained by VM. 
Based on these modules, the symbolic emulation for inter-contract analysis is provided and then the verification of vulnerabilities is executed. 
The details of each process are presented below. 

\subsection{Symbolic Re-entrancy Emulation} 

The goal of this process is to generate a CFG completely via emulation of re-entrancy attacks. 
In particular, the CFManager generates a CFG to represent inter-contract behavior by recording basic blocks with transitions as edges. We call such a CFG an \textit{extended CFG (ECFG)} for convenience. 
In ECFGs, \opcreate\ and \opcall\ instructions are utilized as separators of basic blocks in addition to  
\opjump\ and \opmystop\ instructions. 
For convenience, let contracts to be executed by \opcreate\ and \opcall\ be callers, and let contracts to be created or called be callees.
When \opcreate\ appears, 
the symbolic execution is transited to its callee contract, i.e., the initialization code. 
Similarly, when \opcall\ appears, 
the execution is transited to a function of its callee contract. 
When \opmystop\ appears within the callee contracts described above, the execution is returned to the caller contract. 
These transitions of contracts are managed by VM. 

We now describe the emulation process by RA in detail. 
For instance, for analysis of the \createbase attacks, VM extracts an initialization code of a contract from variables of \opcreate\  by specifying basic blocks via the Z3 SMT solver, and CFManager transits the execution to those blocks in accordance with branches decided by the Z3 SMT solver. 
By obtaining each block and connecting them, \opcall\  in the initialization code can be identified. 
Moreover, VM can register the callee contract obtained by the initialization code in the contract queue, and thus the whole contract can be symbolically emulated in a recursive manner. 
On the other hand, for analysis of the \crossfunction attacks, a function ID is symbolically executed as a symbolic value, and then Verifier extracts the function ID to be executed by the Z3 SMT solver. 
By giving the ID at the beginning execution and re-execution, combination of any functions is representative.

\subsection{Vulnerability Verification}

\begin{figure}[t]
        \begin{center}
            \includegraphics[width=\textwidth,height=3cm]{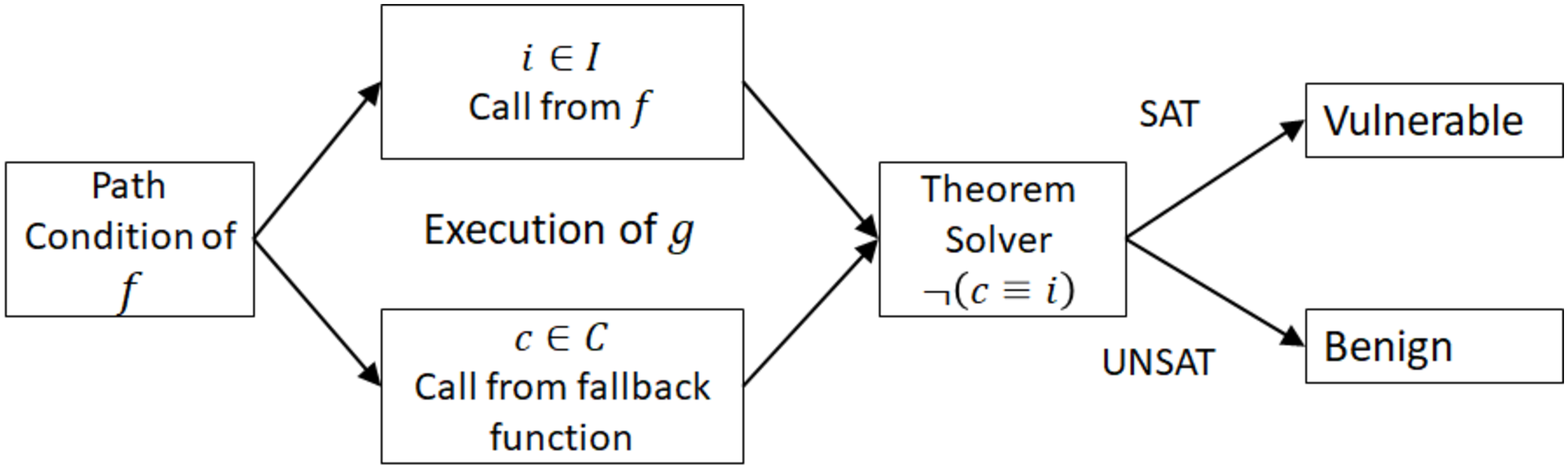}
        \end{center}
        \caption{Process of Vulnerability Verification}
        \label{verification_image}
    \end{figure}

The vulnerability verification is done by Verifier with path conditions obtained from the emulation described in the previous section. 
Let functions to be verified be $f,g$, where $f$ contains a function call during the execution. 
We also denote by $I$ a set of path conditions where $g$ is executed by taking over a result in the execution of $f$, and by $C$ a set of path conditions where $f$ calls a fallback function in a manner that the fallback function calls $g$ and then $f$ is executed again with the result of $g$. The Verifier module can receive $I$ and $C$ from the VM module. 
Then, the Z3 SMT solver verifies whether a program is vulnerable or not as follows: 
\begin{equation}
\exists c \in C , \forall i \in I :  \neg (c \equiv i). \label{verification_property} \end{equation}
The procedure of the vulnerability verification is shown in Fig.~\ref{verification_image}. 
The withdraw process is not executed when $g$ is executed after executing $f$ in a normal way, i.e., executions in $I$. 
On the other hand, the withdraw process is executed due to the re-entrancy on $f$ whereby a fallback function calls $g$, i.e., executions in $C$. 
RA decides that a contract is vulnerable if program behaviors are equivalent in these executions. In particular, RA checks if there is no case where the behaviors are equivalent.
Intuitively, the main difference between these cases is whether instructions following \opcall\ in $f$ are executed before $g$. 
Specifically, $g$ contains a branch that determines whether the withdraw process is executed in accordance with changing states on the blockchain, where path conditions at the end of executions for both cases are different from each other. 

\subsection{Implementation} \label{Implementation}

The main techniques for implementing the symbolic re-entrancy emulation of RA are presented as the technical parts of the implementation below. In particular, we describe how basic blocks are controlled and function IDs are extracted to reduce potential state explosion, which is one of the problems for static analysis tools according to Rodler et al.~\cite{rodler2018sereum}. 
Several techniques for improving the performance are presented as well.
We also plan to release the source codes of the whole implementation publicly via GitHub\footnote{The repository is publicly available. (\url{https://github.com/wanidon/RA})}. 

\textbf{Environment.} 
RA is implemented on Python 3.7 and three modules, i.e., z3, pysha3, and Graphviz. 
The z3 module provides APIs for the use of the Z3 SMT solver on Python. 
The pysha3 module enables the use of the Keccak256 hash function as a SHA-3 module. 
Finally, Graphviz provides support for drawing CFGs. 
We also note that RA is independent of any specific compiler because it uses bytecodes as input.

\textbf{Control of Basic Blocks for CFManager.} 
The method that represents a branch with a conditional jump operation \opjumpi\ is as follows. 
First, RA adopts the depth-first search for finding feasible paths, and the CFManager owns a data structure named \textit{dfs\_stack} that stores candidates of basic blocks to be searched.
Then, the CFManager pops two variables for \opjumpi\  from the dfs\_stack, i.e., a jump address and a condition $c$ for the jump where $c$ is determined by true or false. 
When $c$ is a symbolic value, the z3 module decides whether $\neg c$ is satisfiable. 
If so, there exists a path that does not contain jump. 
Then, a new block is generated by copying the current basic block and incrementing a program counter, and the new block is newly recognized as the basic block. 
Next, the solver decides whether $c$ is satisfiable. 
If so, there exists a path that contains jump. 
Then, the current basic block is copied as a jump block and the program counter is set as the jump address. In the case that there exists a path with jump but a path without jump does not exist, the current basic block is set as a jump block. 
In contrast, if both a path with jump and that without jump exist, then the jump block itself is stored in the dfs\_stack.

\textbf{Extraction of Function ID for Verifier.} 
Decision of a function ID for Verifier is implemented as follows. 
First, a symbolic value $function\_id$ is assigned with the most significant four bytes for any transaction. 
The value is loaded onto a contract to be analyzed, and hence paths are branched for each function by comparing them with actual function IDs for the contract. 
Then, the function ID can be obtained because a solution satisfying $function\_id$ is obtained by deciding the satisfiability of path conditions for all paths with the z3 module. 
If a \opcall\ instruction appears during the execution, \textsf{CALLABLE} is recorded as a state of a basic block. 
The state of the basic block is inherited by all the descendant paths. 
In extracting an actual function ID, ID of a function with function call is recorded if the end state of a basic block is \textsf{CALLABLE}. 

\textbf{Speed-up Techniques}. 
The entire performance of RA was improved by using four optimization techniques. 
First, RA can verify combinations of function calls described in Equation~(\ref{verification_property}) in parallel. 
Second, a constraint is given on push to a stack. A \opcall\ instruction normally requires push to a stack in accordance with success of a function call, i.e., 0 or 1. 
In contrast, a \reentrancy attack is always executed after the success of a function call. Thus, RA always pushes 1, i.e., the success of the function call, to a stack as assumed in the success of \opcall\ instruction.
This enables RA to reduce the number of path conditions as a speed-up technique. 
Third, because \opcall\ is required to always succeed, a constraint, i.e., a balance for any contract account is a positive value, is given on path conditions only at the end states. 
The computational complexity can be reduced drastically by giving the constraint at the end states instead of at each change of the balance. 
Surprisingly, for analysis of several contracts, the computational performance becomes ten times faster by using this constraint although we omit the details due to space limitation.
Finally, RA stops to analyze paths for some stop instruction without the rollback process. 
In particular, a stop instruction called \oprevert\ contains the rollback about the blockchain related to a given transaction. 
However, to the best of our knowledge, \reentrancy attacks never occur in this case, thus RA no longer analyzes the remaining parts of such paths.

\section{Evaluation} \label{Evaluation}

In this section, we show that RA can analyze inter-contract control flows precisely in comparison with execution of Oyente~\cite{luu2016making}. Comparison with other tools is theoretically discussed in Section~\ref{Discussion}. 
The computational performance of RA for analysis is also shown. 

\subsection{Case Studies}

As case studies, we test RA with reference implementations of the re-entrancy attacks by Rodler et al.~\cite{rodler2018sereum} and known re-entrancy vulnerabilities~\cite{solidityDoc,knownattacks}. 
The goal of these studies is to clarify the ability of RA to perform inter-contract analysis. 
The codes of the reference implementations of the \reentrancy attacks~\cite{rodler2018sereum} are publicly available\footnote{\url{https://github.com/uni-due-syssec/eth-reentrancy-attack-patterns}}. 

\textbf{Creation of Contracts.} 
On a CFG output by RA for a contract created via \opcreate, 
the execution is transited to a callee contract on the basic block (denoted by the red box), and the execution is returned on the basic block (denoted by the green box). In contrast, Oyente cannot transit execution into the initialization code, i.e., a created contract, in a symbolic manner. 
The remaining parts of the CFG are identical to a CFG output from Oyente. Consequently, RA can handle a function call by extracting the initialization codes on \opcreate\ instructions. 



On a CFG output by RA on analysis of the reference implementation of the \createbase attack, the execution is transited/returned to/from a callee contract on the basic block. 
This confirms that, unlike Oyente, RA provides analysis of \createbase attacks in an inter-contract fashion.


\textbf{Call of Contracts.} 
Suppose that an adversary utilizes a contract with only \opcall\ instructions. 
In particular, because there are two basic blocks, 
two function calls and two transitions to the caller are identified according to the analysis results of RA. 
These results confirm that RA provides analysis of \crossfunction attacks in an inter-contract fashion. 

\subsection{Vulnerability Analysis}

The performance of RA is evaluated by using it to analyze the vulnerability of a number of deployed contracts to \reentrancy attacks. 
In particular, the target contracts included in the evaluation are the \textit{Fund} contract in the official document of Solidity~\cite{solidityDoc}, a contract named \textit{KnownReentrancy} obtained from the code of ``Reentrancy on a Single Function" published on the webpage of ``Ethereum Smart Contract Best Practices"~\cite{knownattacks}, 
the \textit{Bank} contract based on \createbase attack~\cite{rodler2018sereum}, 
the \textit{Token} contract based on \crossfunction attack~\cite{rodler2018sereum}, and a contract named \textit{KnownCrossFunction} obtained from the code of ``Cross-function Reentrancy" published on ``Ethereum Smart Contract Best Practices". 
Both Fund and KnownReentrancy contain a single contract. 
Bank contains three contracts, two of which are benign, i.e., not vulnerable. 
Token contains twelve contracts, eleven of which are benign and only one contract is vulnerable. 
Finally, KnownCrossFunction contain one vulnerable contract and one benign contract because the cross-function re-entrancy attack needs ``cross" calls between different contracts. 
Several contracts include multiple functions and hence analysis of re-entrancy attacks becomes complicated due to combinations of function calls as described in Section~\ref{Technical Difficulty}. 
In doing so, the goal of this evaluation is to determine whether RA can precisely identify both vulnerable contracts and benign contracts. 

The results are shown in Table~\ref{test-ra}. 
According to the table, RA can precisely verify all the existing vulnerabilities without false positives and false negatives. 
In other words, in addition to being able to analyze contracts that are vulnerable to \reentrancy attacks, 
RA can also analyze contracts that are not vulnerable, as confirmed by the true negative rate.
This precise evaluation for verification of vulnerabilities was obtained by the use of the Z3 SMT solver with path conditions obtained from symbolic execution. 

\begin{table}[tb]
    \centering
    \caption{Results of Analysis of Contracts Vulnerable to Re-Entrancy Attacks: The Benign Functions and Vulnerable Functions columns represent the number of benign functions and vulnerable functions in the contract, respectively. The true positive rate is denoted by TPR, false positive rate by FPR, true negative rate by TNR, and false negative rate by FNR. }
    \label{test-ra}
    \begin{tabular}{c|c|c|c|c|c|c|c}\hline 
    Contract & \shortstack{Code \\Length} [Byte] & \shortstack{Benign \\ Functions} & \shortstack{Vulnerable \\ Functions} & 
   TPR [\% ] &  FPR [\% ] & TNR [\% ] & FNR [\% ] \\ \hline \hline
    Fund~\cite{solidityDoc} & 356 & 0 & 1 & 100 & 0 & 0 & 0 \\ \hline
    \shortstack{Known\\Reentrancy~\cite{knownattacks}} & 365 & 0 & 1 & 100 & 0 & 0 & 0 \\ \hline
    Bank~\cite{rodler2018sereum} & 1694 & 2 & 1 & 100 & 0 & 100 & 0 \\ \hline
    Token~\cite{rodler2018sereum} & 2516 & 11 & 1 & 100 & 0 & 0 & 0 \\ \hline
    \shortstack{KnownCross\\Function~\cite{knownattacks}} & 680 & 1 & 1 & 100 & 0 & 100 & 0 \\ \hline
    \end{tabular}
\end{table}

\subsection{Computational Performance}

We now present the computational performance of RA for analysis of \reentrancy attacks. 
The performance was measured by utilizing the time.perf\_counter function as an average of ten executions. 
The environment for measurement is as follows: iMac 21.5-inch, 2017 with 3.6GHz Intel Core i7 processor, 32GB memory, and Radeon Pro 560 4GB as GPU. 
The measurement was done by implementing parallel-processing. 
As can be seen in Table~\ref{compcost-ra}, the computational time for analysis becomes larger in proportion to the number of function calls. 
Surprisingly, analysis of Bank is fast even with its code length because Bank is based on \createbase attack, i.e., the majority of its codes to create a contract and the function call are restricted to call the created contract. 
In contrast, Token is based on \crossfunction attack and its code length and variation of function calls are large. Nevertheless, RA was still able to analyze Token within reasonable time.

\begin{table}[tb]
    \centering
    \caption{Computational Time of RA for Analysis: The Combinations of Functions column represents the number of combinations of function calls in the contract.}
    \label{compcost-ra}
    \begin{tabular}{c|c|c|c}\hline 
    Contract & \shortstack{Code Length} [Byte] & \shortstack{Combinations of Functions} & Time [sec]  \\ \hline \hline
    Fund~\cite{solidityDoc} & 356 & 1 & 20.750 \\ \hline
    KnownReentrancy~\cite{knownattacks} & 365 & 1 & 23.003 \\ \hline
    Bank~\cite{rodler2018sereum} & 1694 & 3 & 47.579 \\ \hline
    Token~\cite{rodler2018sereum} & 2516 & 12 & 519.676 \\ \hline
    KnownCrossFunction~\cite{knownattacks} & 680 & 2 & 37.705 \\ \hline
    \end{tabular}
\end{table}

\section{Discussion}  \label{Discussion}

In this section, several considerations such as comparison to other existing tools and the current limitations of RA are discussed. 


\textbf{Comparison to Other Analysis Tools.} 
Table~\ref{comparison table} shows a comparison of different tools in terms of their ability to detect re-entrancy attacks discovered by Rodler et al.~\cite{rodler2018sereum}. 
Oyente~\cite{luu2016making} and Securify~\cite{tsankov2018securify} cannot detect such re-entrancy attacks. In particular, Oyente fails to detect the vulnerabilities, while Securify produces false alerts due to its conservative policy. 
Sereum and {\AE}GIS can potentially detect the vulnerabilities, but they are based on dynamic analysis. 
Finally, Annotary is a static analysis tool that can deal with \opcreate\ and \opcall\ instructions. Thus, it can potentially detect the re-entrancy attacks, but its paper did not discuss re-entrancy attacks. Moreover, Annotary's analysis target are Solidity codes and thus it may experience difficulties in analyzing deployed contracts.
\begin{table}[tb]
    \centering
    \caption{Comparison of RA and Other Analysis Tools: $\bullet$ indicates the tool can detect the attack, $\circ$ indicates the tool cannot detect the attack, and $\oslash$ indicates the tool can potentially detect the attack but did not provide a discussion in its paper.}
    \label{comparison table}
    \begin{tabular}{c|c|c|c|c}\hline 
    Tool & \shortstack{Static Analysis} & \shortstack{Cross-Function} & \shortstack{Create-Based} & Analysis Target \\ \hline \hline
    Oyente\cite{luu2016making} & Static & $\circ$ & $\circ$  & EVM \\ \hline
    Securify~\cite{tsankov2018securify} & Static & $\circ$ & $\circ$ & EVM \\ \hline
    Annotary~\cite{Weiss2019} & Static & $\oslash$ & $\oslash$ & Solidity \\ \hline
    Sereum~\cite{rodler2018sereum} & Dynamic & $\bullet$ & $\bullet$ & EVM \\ \hline
    {\AE}GIS~\cite{ferreira2020aegis} & Dyanmic & $\bullet$ & $\oslash$ & EVM \\ \hline
    RA & Static & $\bullet$ & $\bullet$ & EVM \\ \hline
    \end{tabular}
\end{table}

\textbf{Computational Complexity for Vulnerability Verification.} 
Let the number of functions with function call be $F_n$, the number of those paths be $F_p$, the number of any function be $G_n$, and the number of those paths be $G_p$. 
Here, the number of functions to be executed in the first step is $F_n$ and its resulting number of the end state for each execution is $F_p$. 
Then, in the second step, $G_n$ functions are executed in proportion to the number of the end states, and the number of the end states of $G_n$ is $G_p$. 
Consequently, the order complexity to obtain path conditions at the end of process is $O(F_nF_pG_nG_p)$. By parallelizing the process, the complexity is $O(G_nG_p)$ because all combinations of function calls can be parallelized. 
To compute path conditions using Equation~(\ref{verification_property}), RA decides whether two sets, i.e., $C$ and $I$, of the path conditions obtained from combinations of one function have an identity relation between $i\in I$ and $c \in C$. 
In doing so, the order complexity is $O(F_nF_pG_{n}^2G_{p}^2)$, which is smaller than the complexity to obtain path conditions at the end states. 
The execution time for analysis is polynomial time. RA is expected to be utilized for analysis of contracts developed by users in realistic time as shown in Section~\ref{Discussion}. 


\textbf{Extension to Analysis of Delegated Re-Entrancy Attacks.} 
Rodler et al.~\cite{rodler2018sereum} presented delegated re-entrancy attacks where a contract invokes another contract as a library within instructions that utilize contracts as an external library, e.g., \opdelegatecall\  or \opcallcode. 
These instructions are currently not implemented in RA, but RA can be extended to analyze delegated re-entrancy attacks by introducing these instructions. 
The main technical difficulty in analysis of delegated re-entrancy attacks is that it is unknown which library contract will be used~\cite{rodler2018sereum}. This difficulty is also a problem in analyses of create-based and cross-function re-entrancy attacks, which have been overcome already in RA. 

\textbf{Limitations.} 
The current implementation of RA has three limitations. 
First, analysis of gas is not considered. Thus, contracts that restrict gas consumption may not be analyzed precisely. Second, a case where multiple contracts are tightly coupled with each other, i.e., more than two contracts are strongly interdependent, is out of the scope of RA. 
Finally, bytecodes of contracts cannot include symbolic values. Accordingly, a case where created contracts are different for each execution is not considered. 
Improving the points described above is our ongoing work.

\section{Related Works} \label{Related Work} 

In this section, we recall early literature on security analysis of Ethereum smart contracts in terms of symbolic execution and formal methods as static analysis. Then, we describe several multidisciplinary approaches proposed in the past few years as additional related works. 
Interested readers are advised to read the survey paper~\cite{di2019survey} for details on EVM analysis.


\textbf{Symbolic Execution.} 
Symbolic execution of Ethereum smart contracts was originally started by Oyente~\cite{luu2016making}. 
Although there are many subsequent works~\cite{nikolic2018finding,torres2018osiris,liu2018s,chen2017under}, these works do not perform inter-contract analysis. 
Furthermore, analysis of \reentrancy attacks is often heuristic and thus Oyente often produces many false positives. 
Extensions of Oyente that support readability of outputs from symbolic execution have been proposed~\cite{mossberg2019manticore,mueller2018smashing}, and the usability of RA can be potentially be improved in a similar way. 

The closest work to RA is Annotary~\cite{Weiss2019}, which can analyze inter-contract behavior via both symbolic executions of EVM bytecodes and the Z3 SMT solver. The major difference of Annotary from RA is that Annotary mainly targets analysis of Solidity codes. 
In other words, RA mainly checks if deployed contracts are secure or not, while Annotary supports developers in implementing secure codes. 
Although the authors of Annotary did not consider the new attacks by Rodler et al.~\cite{rodler2018sereum}, we consider their idea and work to be elegant nonetheless. 

ETHBMC~\cite{ETHBMC} and VerX~\cite{permenev2020verx} are recent state-of-the-art works that verify properties of Ethereum. VerX is similar to Annotary in terms of taking Solidity codes as input and dealing with external contracts. Similar to Annotary, the new attacks by Rodler et al.~\cite{rodler2018sereum} were not considered in VerX. On the other hand, ETHBMC takes EVM for symbolic executions and its motivation is rather close to that of RA. However, ETHBMC mainly focuses on parity vulnerability. 

\textbf{Formal Methods.} 
Formal verification of EVM was motivated by Bhargavan et al.~\cite{bhargavan2016formal}, and EVM was correctly formalized as KEVM~\cite{hildenbrandt2018kevm}. 
However, verification of the security is challenging in general, and the existing works~\cite{tsankov2018securify,kalra2018zeus} do not provide support for inter-contract analysis, which is the main target of our work. 
As more theoretical approach, Grishchenko et al.~\cite{grishchenko2018semantic} formalized several attacks, including \reentrancy attacks, as well as formalization of EVM. Their formalization inspired the vulnerability verification of RA.

\textbf{Multidisciplinary Approaches.} 
NeuCheck~\cite{Wiley2019} can rapidly analyze Solidity source codes by extracting a syntax tree on a cross-platform environment. TokenScope~\cite{chen2019tokenscope} can detect vulnerabilities by identifying tokens that have a different specification from the ERC20 token. 
SMARTSHIELD~\cite{smartshield} is a bytecode rectification system that fixes security-related bugs automatically. 
Qian et al.~\cite{QL+20} presented an automated re-entrancy detection framework based on machine learning. 
The techniques in these multidisciplinary works can be used to potentially improve RA. 
Finally, ILF~\cite{he2019learning} and ContractWard~\cite{ContractWard} use a combination of machine learning and symbolic execution to improve testing coverage. 
These works are expected to have improved performance if they deploy RA as a building block. 


\section{Conclusion}  \label{Conclusion} 

In this paper, we introduced RA, a static analysis tool that provides inter-contract analysis of the EVM bytecodes of Ethereum smart contracts to detect vulnerabilities to state-of-the-art re-entrancy attacks~\cite{rodler2018sereum}. 
Using RA, analysts do not need to have prior knowledge of \reentrancy attacks to detect them. To create RA, 
we designed modules that represent inter-contract CFGs by the symbolic re-entrancy emulation and the vulnerability verification with the Z3 SMT solver to verify the re-entrancy vulnerability of deployed contracts. 
We also conducted experiments on deployed contracts and confirmed the performance of RA by precisely identifying combinations of contracts with and without vulnerabilities. 
The aforementioned performance could be obtained by virtue of high-level combination of the symbolic execution and the Z3 SMT solver. 


As future work, we plan to extend RA for a case in which a fraction of contracts to be verified and external contracts becomes many-to-many. If successful, RA will be able to deal with more complicated attacks that can be proposed in the future. 
Moreover, we plan to design capabilities that check for vulnerabilities aside from re-entrancy, such as time-dependent vulnerability which utilizes a timestamp for each block. 
We believe that the inter-contract analysis capability of RA can potentially identify such a vulnerability. 
Finally, we will also try to improve the computational performance of RA by reusing parts of the computations for vulnerability verification. 

\subsubsection{Code Availability}
The source code of RA is available in GitHub: \\
\url{https://github.com/wanidon/RA}.

\subsubsection{Acknowledgement} 
This work was supported in part by the Innovation Platform for Society 5.0 at MEXT, and by Secom Science and Technology Foundation.

\end{document}